\pdfoutput=1
\documentclass[11pt,twoside,a4paper,cmspaper,final,collab]{cms-tdr}
\def\svnVersion{9a1bda7}\def\svnDate{2026/07/13}\def\cmsCernNoTag{CERN-EP-2026-007}\def\cmsCernDate{\today}\def\cmsMessage{Published in Physical Review D as \href{http://dx.doi.org/10.1103/q2mh-7mhj}{\doi{10.1103/q2mh-7mhj}.}}
\begin{document}\cmsNoteHeader{HIG-24-014}

\newcommand{\MGG}{\ensuremath{m_{\PGg\PGg}}\xspace}
\newcommand{\PHI}{\ensuremath{\upphi}\xspace}
\newcommand{\ggPHI}{\ensuremath{\Pg\Pg \to \PHI}\xspace}
\newcommand{\PHIgg}{\ensuremath{\PHI \to \PGg\PGg}\xspace}
\newcommand{\Zee}{\ensuremath{\PZ \to \Pep \Pem}\xspace}
\newcommand{\Zmumug}{\ensuremath{\PZ \to \PGmp \PGmm \PGg}\xspace}
\newcommand{\parta}{\HepParticle{a}{}{}\xspace}
\newcommand{\Pj}{\HepParticle{j}{}{}\xspace}
\newcommand{\CP}{\ensuremath{CP}\xspace}

\newlength{\cmsFigWidth}
\ifthenelse{\boolean{cms@external}}
{ \setlength{\cmsFigWidth}{\columnwidth}}
{ \setlength{\cmsFigWidth}{0.6\textwidth}}

\cmsNoteHeader{HIG-24-014}
\title{Search for a narrow resonance with a mass between 10 and 70 GeV decaying to a pair of photons in proton-proton collisions at \texorpdfstring{$\sqrt{s} = 13\TeV$}{sqrt(s) = 13 TeV}}

\date{\today}

\abstract{
The existence of a new spin-zero particle with a mass below the electroweak scale is predicted by several theoretical models. Searches for resonant production of photon pairs at the LHC are able to probe these models. We present a search for a narrow resonance produced through gluon fusion that decays into a pair of photons with an invariant mass between 10 and 70\GeV, using a proton-proton collision data set from the CMS experiment. 
This data set, corresponding to an integrated luminosity of 54.4\fbinv, was recorded in 2018 at a center-of-mass energy of 13\TeV using a newly introduced diphoton trigger that enabled exploration of the low-mass diphoton spectrum. No significant excess above the expected background is observed.
Upper limits are set on the product of the gluon fusion production cross section and the branching fraction of the diphoton decay of a narrow resonance. An interpretation of these limits within an effective field theory framework for axion-like particles is also provided.
}

\hypersetup{%
pdfauthor={CMS Collaboration},%
pdftitle={Search for a very low mass diphoton resonance using 2018 data},%
pdfsubject={CMS},%
pdfkeywords={CMS, exotica, resonance, diphoton, ALP}} %

\maketitle 

\section{Introduction}
\label{sec1:intro}
The standard model (SM) of particle physics provides an elegant and robust framework for describing fundamental interactions. However, it leaves several profound questions about our universe unresolved, 
such as the existence of dark matter (DM)~\cite{Arbey:2021gdg,Planck:2018vyg} and the vast gap between the gravitational and electroweak (EW) energy scales.
These open questions drive experimental searches for new physics.
An extensive program in dark-sector searches is underway at the CERN LHC, looking for the production of low-mass (down to the \GeV scale) and high-mass (up to the multi-TeV regime) resonances across various final states with proton-proton (pp) and heavy ion collisions.
Two-body decays of resonances have played a crucial role in the history of particle physics, leading to landmark discoveries across a wide range of masses, from the $\PJGy$~\cite{JPsi_Aubert:1974js,JPsi_Augustin:1974xw} and $\ensuremath{\Upsilon}$~\cite{Upsilon_Herb:1977ek} mesons at the GeV scale, to the $\PW$ and $\PZ$ bosons~\cite{W_UA1:1983emi,Z_UA1:1983ugx}, the Higgs boson~\cite{Aad:2012tfa,Chatrchyan:2012xdj}, and the top quark~\cite{Top_CDF:1995wbb,Top_D0:1995jca} at the EW scale. 
In recent years, sensitivity to a sub-EW mass range and rare phenomena has been significantly enhanced at the LHC, driven by trigger developments and the exploration of new analysis techniques. 

The discovery of the SM Higgs ($\PH$) boson has brought spin-zero particles into focus as key players in the search for physics beyond the SM (BSM), with scalar and pseudoscalar particles emerging as potential messengers of new physics. 
The fundamental nature of the $\PH$ boson remains unresolved, as it could either be an elementary particle or possess a composite structure. The apparent fine tuning associated with the EW hierarchy~\cite{Gildener:1976ai,Susskind:1978ms,SMITH198138} may simply reflect an intrinsic feature of nature, or it may point to a deeper dynamical mechanism. In the latter case, additional symmetries are often explored as possible explanations for the emergence of a low-mass Higgs-like particle with spin zero.
Numerous generalizations of the SM predict the existence of additional (pseudo)scalar states, such as two-Higgs-doublet models~\cite{Celis_2013,Coleppa:2013dya,Chang:2013ona,Cacciapaglia2016}, the next-to-minimal supersymmetric SM~\cite{Dine:1981rt,Drees:1988fc,King:1995vk,Franke:1995tc,Maniatis:2009re,Ellwanger:2009dp,Ellwanger:2015uaz,1674-1137-38-7-073101,Guchait:2016pes,PhysRevD.39.844,Tao_2018}, Georgi--Machacek models~\cite{GEORGI1985463,PhysRevD.92.075011,PhysRevD.103.095010,Wang_2022}, and composite Higgs models~\cite{Contino:2010rs}. This search is designed to minimize the model-dependent assumptions, ensuring broad applicability to BSM physics scenarios. 
In this paper, we interpret our results in the context of axion-like particles (ALPs), which are pseudo-Nambu--Goldstone bosons arising from the spontaneous breaking of approximate global symmetries~\cite{Marsh:2015xka}. Unlike the QCD axion, which represents a specific realization introduced to solve the strong charge-parity (\textit{CP}) problem in quantum chromodynamics (QCD)~\cite{Peccei:1977hh,Abel:2020gbr}, the mass and couplings of ALPs are generally independent parameters. This allows for the possibility of resonances at the GeV--TeV scale, well within the kinematic reach of the LHC. Such ALPs can couple to photons and other SM particles and appear in a variety of theoretical scenarios. Depending on the framework, ALPs may also play a role in DM or cosmology.

So far, most searches for ALPs~\cite{Mariotti:2017vtv} have focused on their couplings to photons, gluons, and/or quarks, as these interactions dominate at the LHC scale of energies and provide key astrophysical and cosmological constraints for very light ALPs. Current observational probes from supernovae, stellar cooling, and beam dump experiments are
typically sensitive to masses ranging from sub-eV scales up to the \GeV range, while collider searches extend this reach from the \GeV scale up to a few \TeV. 
Previous searches for BSM physics at CMS have explored low-mass resonances in both dijet and diphoton final states, albeit with limited sensitivity in the low-mass regime. A dijet search~\cite{CMS:2017dcz} was limited to invariant masses above 50\GeV due to trigger selection requirements. Similarly, earlier CMS searches for diphoton resonances targeted masses above 70\GeV~\cite{CMS:2024yhz}, limiting sensitivity to ALP signals in the lower-mass phase space. Access to this unexplored phase space became feasible in CMS only in 2018, when a new diphoton trigger with asymmetric transverse momentum (\pt) thresholds of 30 and 18\GeV on the two photons - without any minimum invariant mass requirement, previously set at 55\GeV due to inheriting the requirements for the SM $\PH$ boson search - was deployed.

\begin{figure}[!htb]
\centering
\includegraphics[width=\cmsFigWidth]{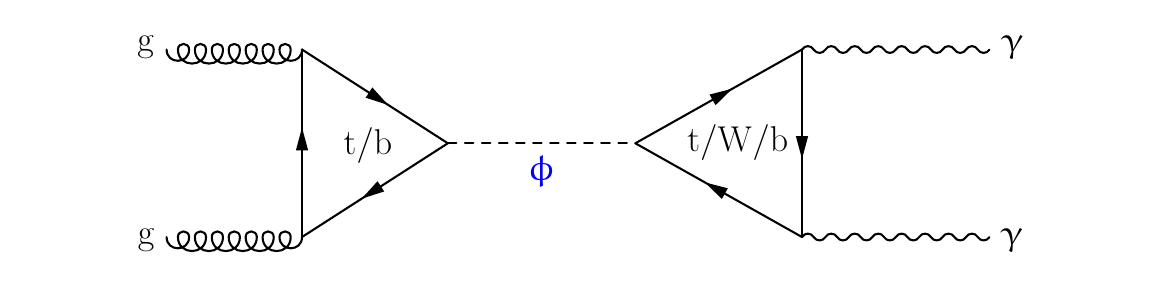}
\caption{Feynman diagram of a generic Higgs-like diphoton resonance $\PHI$ produced via gluon fusion decaying into a pair of photons.}
\label{fig:diagram}
\end{figure}

The ATLAS Collaboration performed a similar search in the diphoton spectrum between 10 and 110\GeV using the entire data set collected during 2016--2018~\cite{ATLAS-CONF-2022-018,ATLAS:2024bjr}.
The constraints on the production of ALPs and their couplings with photons and gluons have been extensively discussed in Ref.~\cite{Mariotti:2017vtv}, which explores the potential for probing the phase space of low-mass diphoton resonances in LHC searches at 7, 8, and 14\TeV, where significant ALP production is anticipated in pp collisions. 
The present search investigates the production of a Higgs-like diphoton resonance, denoted by $\PHI$, with a mass as low as 10\GeV, focusing on the gluon-fusion production mode (Fig.~\ref{fig:diagram}). The resonance is assumed to have the same spin and \CP quantum numbers as the SM $\PH$ boson and couplings to SM particles proportional to their masses, such that interactions with heavy fermions dominate.
Gluon-fusion production represents the dominant process at the LHC \cite{Djouadi:2005gi,deFlorian:2016spz}, particularly at low mass, due to the large coupling of a Higgs-like resonance to top quarks in the loop and the high abundance of gluons in the proton. For the same reason, the decay $\PHIgg$ proceeds predominantly through loops of heavy charged particles, yielding an experimentally clean final state.
It is chosen to keep the search as inclusive as possible in order to simplify the interpretation of exclusion limits within any of the previously mentioned models.

The scope of this study presents unique challenges compared to previous diphoton searches, primarily due to the presence of highly Lorentz-boosted photon pairs with transverse momenta close to the trigger thresholds. Boosted signal topologies dominate in the lower part of the diphoton invariant mass ($\MGG$) spectrum below the threshold region of the trigger (at around 50\GeV), hereafter referred to as the trigger turn-on, whereas less boosted topologies become increasingly accessible and start to dominate at higher masses. At a mass of about 10\GeV and for photons passing the trigger thresholds defined above, the angular separation between the two photons is comparable to the isolation cone used to select photons, making it challenging to extend the search to lower masses. At even lower masses, the highly boosted topology causes the two photons to merge into a single cluster in the electromagnetic calorimeter (ECAL), making standard photon reconstruction ineffective and requiring a dedicated algorithm to resolve such collimated decays~\cite{CMS:2022wjj}. For a 10\GeV signal mass hypothesis, approximately 2\% of events exhibit an angular separation between the two photons smaller than 0.3, indicating that highly collimated photon configurations, which could affect the isolation efficiency, represent only a small fraction of the signal in the considered regime.
To optimize the sensitivity of the search in this challenging region of the diphoton spectrum, a multivariate analysis (MVA) technique based on a neural network (NN) is employed for the event selection, leading to an enhanced signal-to-background ratio and improved mass resolution.
For a signal mass hypothesis of 30\GeV, this selection achieves nearly a factor of four improvement in the signal-to-background ratio compared to the previous MVA strategy, for the same number of background events.

The search targets narrow resonances in the diphoton invariant mass range 10--70\GeV, where the detector mass resolution is approximately 1.5\% of the resonance mass, ensuring sensitivity to signals with widths much smaller than the resolution. The diphoton invariant mass spectrum is modeled with analytic functions to describe both signal and backgrounds, maximizing sensitivity to potential new resonances while accounting for the complex background structure in data. Results of the search are expressed in terms of the product of the inclusive gluon-fusion production cross section and diphoton branching fraction of a narrow and prompt resonance, denoted by ${\sigma(\ggPHI)}{\mathcal{B}(\PHIgg)}$. The use of NN selection optimizes the search for this production mechanism and other production mechanisms are not considered.

The structure of the paper is outlined as follows.
A brief description of the CMS detector and the event reconstruction is given in Section~\ref{sec2:cms}, with a focus on the ECAL and photon reconstruction. The data sets and simulated samples are detailed in Section~\ref{sec3:datasets}, followed by a discussion of the diphoton event selection and signal region definition in Section~\ref{sec4:eventSelCat}, which discusses MVA techniques. 
Sections~\ref{sec5:modelling} and~\ref{sec6:syst} describe the signal and background modeling employed in the statistical analysis as well as the main sources of systematic uncertainty, respectively. The final results and their recasting in the context of ALP models are described in Section~\ref{sec7:results}. The paper is summarized in Section~\ref{sec8:summary}.
The results of this analysis, presented as constraints on the production of a narrow, promptly decaying diphoton resonance and their extrapolation to the ALP phase space, are provided in the HEPData record~\cite{hepdata}. 

\section{The CMS detector and event reconstruction}
\label{sec2:cms}
The central feature of the CMS detector is a superconducting solenoid of 6\unit{m} internal diameter providing a uniform magnetic field of 3.8\unit{T}. Within the solenoid volume are a silicon pixel and strip tracker, a lead tungstate crystal ECAL, and a brass and scintillator hadron calorimeter (HCAL), each composed of a barrel and two endcap sections. Forward calorimeters extend the pseudorapidity ($\eta$) coverage provided by the barrel and endcap detectors. 
Muons are reconstructed using gas-ionization detectors interleaved with the layers of the steel flux-return yoke outside the solenoid.
A significantly improved event description is achieved by correlating the fundamental elements from all detector layers, such as tracks and clusters, to identify final-state particles. 
Further details about the ECAL detector, which plays a central role in reconstructing the final state targeted in this work, as well as the photon reconstruction process, are presented in the following paragraph.
A two-tiered trigger system is used to select events of interest for the analysis. 
The first level of the CMS trigger system uses information from the calorimeters and muon detectors to select the most interesting events in a fixed time interval
of approximately 4\unit{$\mu$s}~\cite{CMS:2020cmk}. The high-level trigger processor farm further decreases the event rate from approximately 100\unit{kHz} to a few \unit{kHz} before data storage~\cite{CMS:2016ngn,Hayrapetyan_2024}.
A comprehensive overview of the CMS detector, including details on the coordinate system and relevant kinematic variables, can be found in Refs.~\cite{Chatrchyan:2008zzk,CMS:2023gfb}.

\vspace{0.5cm}\paragraph*{The ECAL and photon reconstruction}
The energy of photons, which play a central role in this search, is measured in the ECAL. The ECAL is a hermetic, homogeneous calorimeter made of 61\,200 lead tungstate crystals, mounted in a central barrel (EB), and 7\,324 crystals in each of the two endcaps (EE). The EB covers $\abs{\eta}<1.479$, while the EEs cover $1.479<\abs{\eta}<3.0$. Avalanche photodiodes are used as photodetectors in the barrel and vacuum phototriodes are used in the endcaps. In the pseudorapidity region $1.65<\abs{\eta}<2.6$, a three radiation length preshower detector with two orthogonal layers of silicon strips is placed in front of the endcap crystals. 
In the EB (EE), an energy resolution of approximately 1.0\% (2.5\%) is achieved for unconverted or late-converting photons in the tens of GeV energy range. For other photons, the resolution is around 1.3\% up to $\abs{\eta}=1$, increasing to about 2.5\% at $\abs{\eta}=1.4$, and ranging 3--4\% in the EE~\cite{CMS:EGM-14-001}.
The reconstruction process starts by identifying seed crystals in the ECAL as local energy maxima. Around each seed, neighboring crystals are aggregated with the seed crystal if they have an energy above a specific threshold. These clusters are subsequently merged into "superclusters," with the preshower energy included when applicable~\cite{CMS:EGM-14-001, Chatrchyan:2013dga}.
Photon energy is computed from the calibrated crystal energies, with corrections for shower containment and conversion losses using an MVA regression~\cite{CMS:EGM-14-001,Sirunyan:2020ycc}.
This regression also provides a per photon energy resolution estimate for every event. 
After applying the regression correction, any remaining differences between data and simulation are addressed using $\PZ\to\EE$ events, where the track of electrons is ignored and the energy deposit in the ECAL from the electron is interpreted to a photon. An additional set of corrections is applied to align the mean of the dielectron mass spectrum in data with the expected value from simulation. A smearing is also applied to the ECAL hit energies in simulation to reproduce the resolution observed in data. Further details on this procedure can be found in Ref.~\cite{CMS:2020xrn}.

\section{Data sets and simulated samples}
\label{sec3:datasets}
The analyzed data sample corresponds to an integrated luminosity of 54.4\fbinv collected in pp collisions at a center-of-mass energy of 13\TeV during 2018. A diphoton trigger with \pt thresholds of 30 and 18\GeV on the two leading photons was implemented in 2018, removing the previously applied minimum mass requirement of 55\GeV to ensure access to the phase space relevant to this analysis. The trigger efficiency is measured with the ``tag-and-probe'' technique~\cite{tagNprobe} using $\PZ\to\EE$ events. 
These efficiencies derived from data are then applied as corrections to the simulated signal events, and the associated statistical and systematic uncertainties are propagated to the expected signal yields.
More details on the trigger and its efficiency can be found in Ref.~\cite{CMS:2024yhz}.
Throughout the analysis, the $\PZ$ resonance serves as a well-understood, high event count benchmark for robust object validation across the full analysis phase space. Since the same object requirements are applied, only the resonance boost is expected to differ when extrapolating to lower masses. Potential systematic uncertainties arising from differences between the $\PZ$ boson and resonances at lower mass hypotheses were studied and are described later in Section~\ref{sec6:syst}.

The signal is modeled using simulated events, with several corrections applied to improve the agreement between Monte Carlo (MC) simulation and data. These corrections are derived from control samples, specifically $\Zee$ and $\Zmumug$ events, and applied to the simulated samples to improve and validate the modeling of photon kinematics and selection efficiencies. 
Signal samples are simulated using MC techniques. In this analysis, a SM-like $\PH$ boson, $\PHI$, serves as a proxy for a generic spin-zero diphoton resonance. Events are generated with masses ranging from 10 to 70\GeV in steps of 5\GeV. The simulation of $\PHI$ signals produced via gluon fusion ($\Pg\Pg\PHI$) employs the \MGvATNLO v2.6.5~\cite{Mad5} matrix element generator~\cite{AMCAT} at next-to-leading order (NLO) in perturbative QCD with FxFx merging~\cite{FXFX} and the NNPDF3.1~\cite{Ball:2017nwa} set of parton distribution functions
(PDFs). 
Parton showering and hadronization of the parton-level particles are performed using the \PYTHIA8.205~\cite{Sjostrand:2014zea} interface, with the CP5~\cite{Sirunyan:2019dfx} underlying event tune.

The dominant backgrounds for diphoton signal signatures arise from irreducible prompt diphoton production as well as $\PGg+$jets and multijet events in which one or two jets produced through the strong interaction are misidentified as isolated photons. The former dominates at higher diphoton masses (above the trigger thresholds), with its contribution diminishing as the mass decreases. In contrast, the contribution from $\PGg+$jets and multijet events becomes the leading background in the lower-mass region. An accurate representation of the background sources is essential for the development and optimization of the NN used to select signal region (SR) events, as described in the next Section. The irreducible contribution is modeled from simulation, while the reducible component of the background, dominated by misidentified photons, is modeled from data.
The diphoton background is simulated at leading order (LO) with the \SHERPA v.2.2.11~\cite{SHERPA} generator with matrix-element-plus-parton-shower merging~\cite{Hoeche:2009rj} in the mass range 0--80\GeV. 
It includes tree-level diagrams with up to three additional jets along with box diagrams.
A simulation of $\PGg$+jets and multijet backgrounds, performed using the \PYTHIA generator at LO, is also utilized for auxiliary studies, such as establishing the necessity of estimating the reducible background directly from data.
Control samples, such as Drell--Yan $\Zee$ and $\Zmumug$ processes, used to derive the photon corrections and their systematic uncertainties, are simulated with \MGvATNLO either at LO or NLO.

To account for additional pp interactions occurring in the same or nearby bunch crossings (pileup), the \PYTHIA generator is employed. 
All MC samples include effects of pileup by overlaying simulated minimum bias interactions on the hard scattering event, with the multiplicity distribution matching that in data.
The events are subsequently processed through a full CMS detector simulation using the \GEANTfour toolkit~\cite{Agostinelli:2002hh}.

\section{Event selection}
\label{sec4:eventSelCat}
The analysis selects diphotons with an invariant mass between 10 and 70\GeV.
The identification of the primary vertex from which the two photons originate is performed with a boosted decision tree (BDT), following the approach described in Ref.~\cite{CMS-HIG-16-040}. The BDT output also provides a per-event probability estimate of correctly identifying the primary vertex. The BDT utilizes inputs such as the sum of squared \pt of charged-particle tracks associated with each vertex, along with variables that characterize the momentum balance between the diphoton system and these tracks, accounting for photon conversions. Accurate identification of the diphoton vertex is crucial for achieving optimal diphoton invariant mass resolution. In fact, when the position of the interaction along the beam axis (z) is determined within approximately 1\unit{cm} (which occurs in more than 90\% of events), the mass resolution is predominantly dictated by the photon energy resolution, typically ranging 1--3\%, as discussed earlier. 

A set of offline criteria on the photon objects, which are slightly tighter than those used at the trigger level, is applied to ensure that the trigger is nearly 100\% efficient.  This set, referred to as the photon preselection, includes selections on the photon \pt, the ratio of energy deposited in HCAL to that in ECAL, the shape of the electromagnetic shower, and the isolation of the photon supercluster. 
Each photon candidate in the pair meets specific criteria on its \pt and the ratio of its \pt value to the invariant mass of the diphoton system: $\pt^{\PGg_1(\PGg_2)} >30\,(18)\GeV$ and $\pt^{\PGg_1 (\PGg_2)}/m_{\PGg\PGg}>0.47~(0.28)$, with $\PGg_1\,(\PGg_2)$ referring to the photon candidate in the pair with the higher (lower) \pt value. 
The use of \pt thresholds scaled by $\MGG$ prevents distortions in the diphoton mass spectrum near the upper boundary for masses above 64\GeV, while for lower masses, the selection remains driven by the fixed \pt thresholds of 30/18\GeV. This approach also ensures consistency with the previous low-mass diphoton analysis~\cite{CMS:2024yhz}.
Isolation criteria are applied to both photons by requiring the unclustered energy in ECAL within a cone of radius $\Delta R < 0.3$ around the photon supercluster to be less than 4\GeV, and the scalar sum of track \pt in the same cone to be less than 6\GeV. 
Here, $\Delta R$ is the spatial separation in the $(\eta, \phi)$ plane, defined as $\Delta R = \sqrt{\smash[b]{(\Delta \eta)^2 + (\Delta \phi)^2}}$, where $\Delta \eta$ and $\Delta \phi$ represent the differences in pseudorapidity and azimuthal angle, respectively.
This requirement is particularly important for high-energy photons in boosted topologies, ensuring that they remain sufficiently isolated to allow for the proper application of all available corrections. 
Additional isolation requirements are imposed by limiting the energy deposited in the HCAL by charged particles within a cone of radius $\Delta R < 0.3$ around the photon to be below 20\GeV or less than 30\% of the photon \pt. An electron veto is applied to reject photon candidates geometrically matched to at least two hits in the pixel detector.

Dedicated photon identification criteria were developed in the context of the recently published CMS diphoton search in the mass range 70--
110\GeV~\cite{CMS:2024yhz}, based on a BDT implemented within the \textsc{TMVA} package~\cite{tmva}. This BDT is trained on simulated $\PGg$+jets events, which include both prompt and nonprompt photon candidates, following the procedure described in Ref.~\cite{CMS-HIG-16-040}. It will hereafter be referred to as the photon ID MVA.
The photon ID MVA distinguishes between photons originating directly from hard-scattering processes, referred to as prompt photons, and photons primarily resulting from $\pi^0$ and $\eta$ meson decays within jets, referred to as nonprompt. It provides a continuous estimator for each photon, ranging from $-1$ to $+1$, where nonprompt photons tend to have values near $-1$, while prompt photons cluster near $+1$.
Photon identification inputs, such as shower shape and isolation variables, are corrected in simulation using a chained quantile regression method~\cite{Spyromitros_Xioufis_2016}, based on $\PZ\to\EE$ events. A sequence of individually trained BDTs is used to correct each variable, by incorporating photon kinematics, per-event energy density, and previously corrected variables as inputs. This approach ensures that correlations among the inputs remain consistent and closely resemble those observed in data. 

Each photon candidate is required to have a minimum photon ID MVA score of $-$0.7 in order to minimize the contribution from backgrounds with nonprompt photons, while maintaining good efficiency for prompt diphotons. Simulation studies show that this requirement discards 50\% (35\%) of multijet ($\PGg$+jets) events, while retaining approximately 98\% of prompt diphoton events.
Events with a minimum photon ID MVA score between $-$0.9 and $-$0.7 are instead used to build an orthogonal sideband region (SB), as done in Ref.~\cite{CMS:2020cga}, while the maximum photon ID MVA score is required to be larger than $-$0.7. This SB region, dominated by nonprompt photons, is used to model the combined QCD and $\PGg$+jets components of the background. 
The efficiencies of the photon ID MVA and preselection criteria in the signal simulation are corrected using $\PZ\to\EE$ events.
These corrections are derived as the ratio of efficiencies measured in data to those obtained from simulation with a tag-and-probe technique~\cite{HIG-19-015,CMS:2024yhz}. 
The corresponding statistical and systematic uncertainties are propagated to the expected signal yields.

An MVA technique is employed to classify events as signal or background by assigning higher scores to signal-like events and lower scores to background-like events. This classification is performed using a NN-based diphoton MVA classifier, which relies on 10 input features that are descriptive of the diphoton system. 
These features include: 
\begin{itemize}
    \item the transverse momenta of the photons scaled by the diphoton invariant mass;
    \item their pseudorapidities;
    \item the photon ID MVA scores;
    \item the cosine of the angle between the two photons;
    \item the per-event estimate of the probability of correctly identifying the primary vertex;
    \item the event-by-event estimate of the mass resolution, based on per-event estimates of photon energy resolution and on the longitudinal spread of the interaction region along the beam axis~\cite{HIG-19-015}.
\end{itemize}

In addition, the event mass hypothesis ($m_{\text{hyp}}$) is incorporated as a discrete parameter during training, which aids in capturing the varying behaviors of input variables across the mass range. 
The $m_{\text{hyp}}$ parameter is assigned using the nominal discrete mass value for each signal MC sample. For background events, $m_{\text{hyp}}$ is assigned to be the discrete mass value from the MC signal sample closest to the reconstructed diphoton mass. 
Using $m_{\text{hyp}}$ as a discrete parameter, rather than a continuous one, minimizes sculpting effects in the final $\MGG$ distribution, which serves as the observable in the statistical fit.

The training and validation data sets are constructed with a mixture of signal events from different mass hypotheses (10, 15, 20, 30, 40, 50, 55, 60, and 70\GeV) divided into two statistically independent equal parts. A few simulated signal samples at intermediate mass values (25, 35, 45, and 65\GeV) were not used for training and were used to validate the interpolation model described in the next Section, with no statistically significant discrepancies observed. The background component is represented by diphoton MC events and data events in the SB region.
Since the minimum photon ID MVA score of SB events lies outside the preselection region (as this variable defines the SB selection), but is still used as an input to the NN, a new photon ID MVA score must be generated for SB events. To ensure the SB data accurately emulates the characteristics of events in the preselection region, a probability density function (pdf) derived from $\PGg$+jets MC events is employed to model the distribution of the minimum photon ID MVA within the allowed range. This pdf is used to randomly generate a new minimum photon ID distribution for SB events. Correlations with other variables used in the NN are neglected when generating the new photon ID score.
In addition, the maximum photon MVA ID is reshaped through a reweighting procedure by using the same pdf, in order to match the distribution in the preselection region.
A two-dimensional template fit on the minimum and maximum photon ID of the diphoton MC events and SB events to the data is then performed to ensure that the relative fractions of prompt diphoton and nonprompt photon backgrounds used in training match those observed in data.
Finally, a dedicated reweighting procedure is applied to the SB events as a function of properties of both photons such that the distributions of the NN input variables agree with observed data in the preselection region. This procedure corrects remaining mismodeling of energy resolution, \pt, and $\eta$ for the $\PGg+$jets and multijet backgrounds.

\begin{figure*}[!hbt]
\centering
\includegraphics[width=0.65\textwidth]{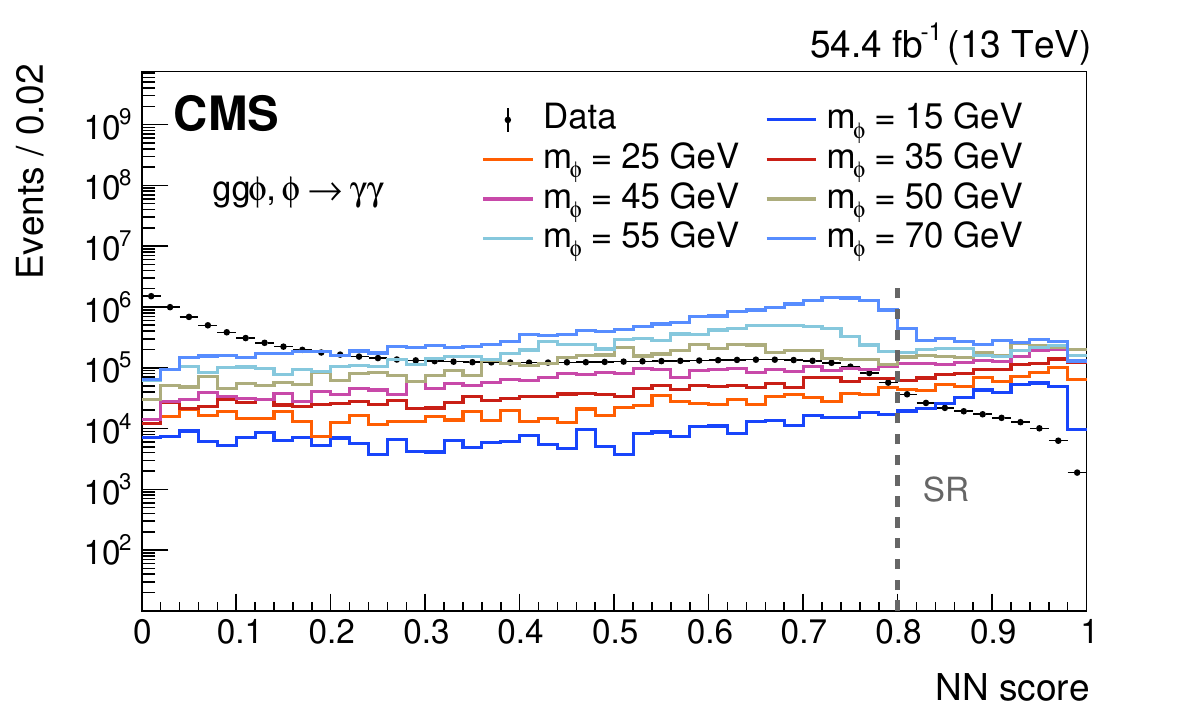}
\caption{Distribution of the NN score for signal events produced via gluon fusion at different mass hypotheses, normalized to 100\unit{nb} for illustration purposes.
The observed distribution of data in the preselection region, serving as a representation of the background events, is shown as black dots. The NN score threshold of 0.8 used to define the SR is indicated by the gray dashed line.
}
\label{fig:pnn}
\end{figure*}

The NN training is based on the \textsc{TensorFlow} package~\cite{tensorflow2015-whitepaper}. The architecture consists of one input layer with eleven neurons (one for each input variable, plus the mass condition), three hidden layers consisting of fifty neurons each, and one output layer that uses the rectified-linear-unit activation function~\cite{Relu:Nair2010}. The Adam optimizer~\cite{Adam:Kingma2014} is used, and the loss is calculated using a weighted cross-entropy function. 
The signal and background event scores are evaluated using the reconstructed $\MGG$ in place of the discrete mass hypothesis adopted in the training. Figure~\ref{fig:pnn} shows the distribution of the NN score for signal events under different mass hypotheses, together with the distribution observed in data after the preselection. 
The NN score distribution observed in data is well reproduced by the background sample used for the training, which consists of a mixture of irreducible prompt diphoton background events as well as $\PGg+$jets and multijet events.
The NN output distributions show a good signal-background separation, with a shoulder appearing between 0.5 and 0.8 corresponding to less boosted events. 
This region corresponds to the trigger turn-on point, where the signal can pass the trigger thresholds without necessarily being boosted, enabling a mixture of boosted and nonboosted diphoton topologies and thus a change in the shape of the NN distribution. 
The NN performance is evaluated in terms of area under curve values at every mass in steps of 5\GeV and is found to be stable (above 80\%) across the full mass range of the search.
An improvement of 4--24\% over the diphoton MVA used in the 70--110\GeV diphoton analysis~\cite{CMS:2024yhz} is achieved, depending on the mass hypothesis. The largest gain is observed in the challenging region below 50\GeV.

The SR is defined using the output of the NN classifier and represents the subset of diphoton events with the highest sensitivity for the search. 
A fixed threshold of 0.8 on the NN score is applied to define the SR. This threshold is chosen to optimize the approximate Asimov significance~\cite{Cowan:2010js} as the figure of merit, while ensuring sufficient event count for both signal and background modeling across the entire mass range. High sensitivity is achieved for all masses studied, and it was checked that including additional NN categories results in negligible improvement. This single criteria tends to favor boosted topologies in the entire mass range and lowers the effect of the trigger turn-on, where nonboosted topologies are prevalent.

\section{Signal and background modeling}
\label{sec5:modelling}
Accurate interpretation of the results requires a reliable modeling of the diphoton signal and background. Simulated diphotons from $\Pg\Pg\PHI$ events, including the aforementioned corrections from data/MC comparisons in $\PZ$ events, are used to determine the product of efficiency and acceptance, as well as the shape of the diphoton spectrum as a function of the mass. 

\begin{figure*}[!hbt]
\centering
\includegraphics[width=0.49\textwidth]{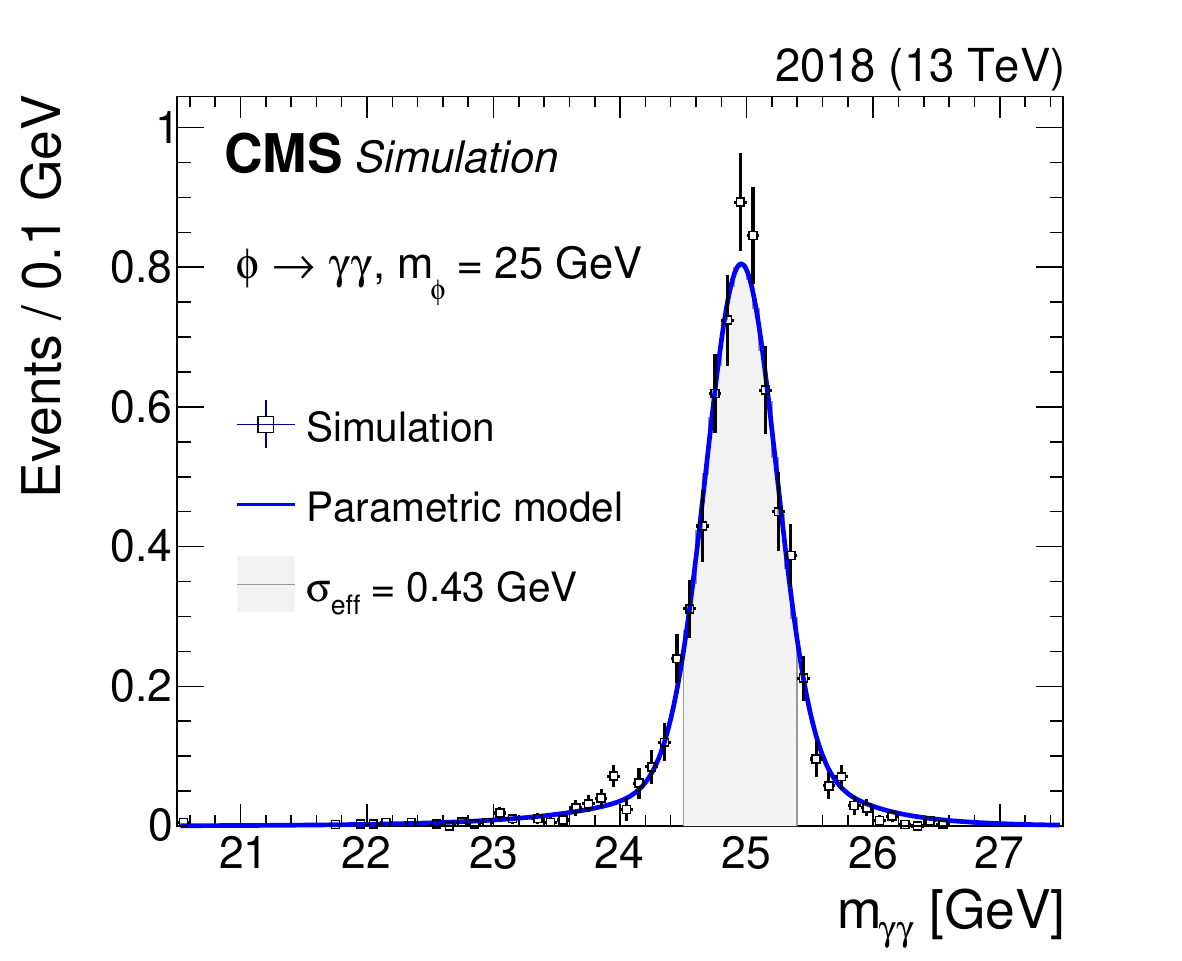}
\includegraphics[width=0.49\textwidth]{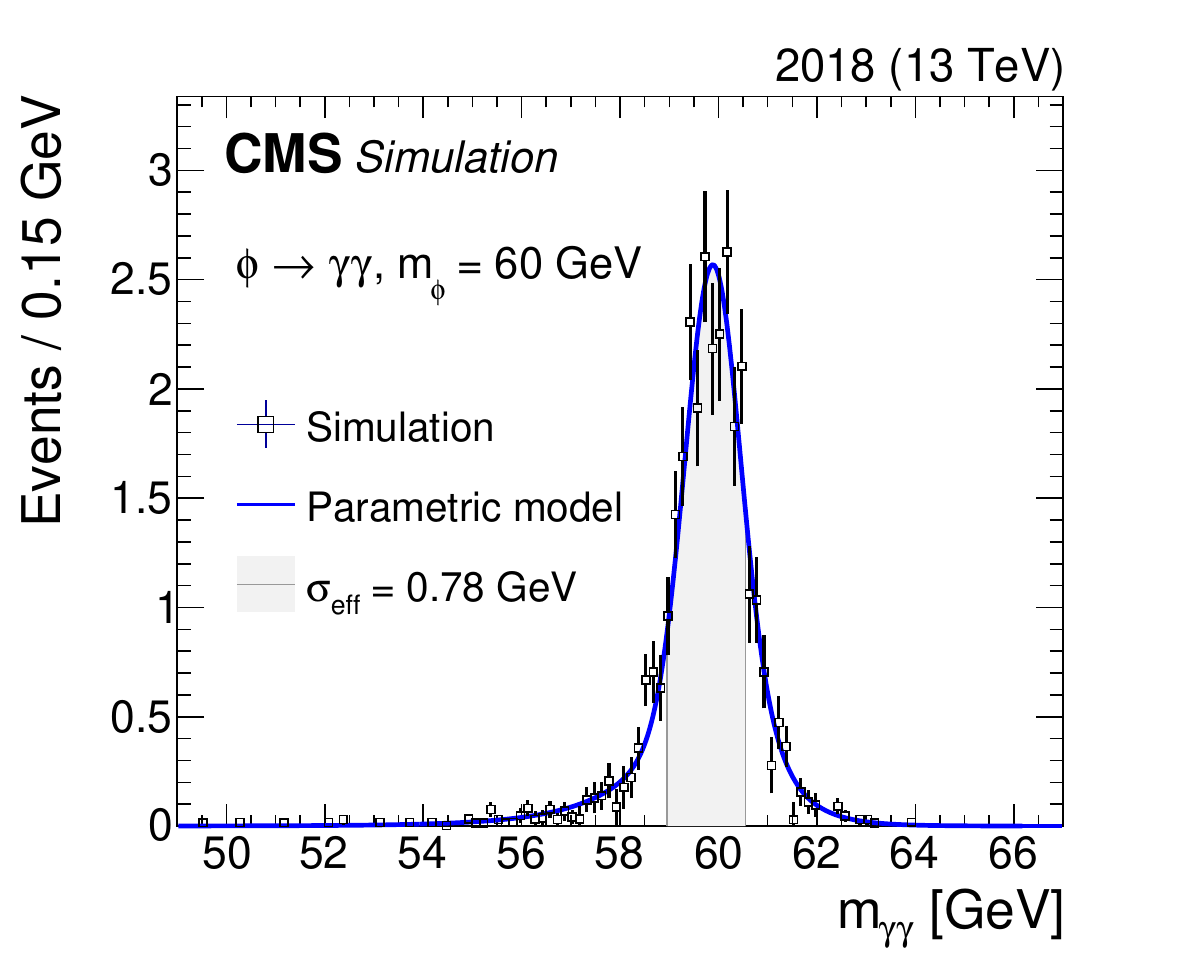}
\caption{The parametric signal model (blue solid line), derived from simulation of Higgs-like diphoton signals (square markers), is evaluated at two mass hypotheses: 25\GeV (left) and 60\GeV (right). The typical experimental resolution for the signal is also displayed in terms of the effective width ($\sigma_{\text{eff}}$), corresponding to the shaded region. The $\sigma_{\text{eff}}$ is defined as half of the smallest interval that contains 68\% of the total probability density distribution.
}
\label{fig:sigModel}
\end{figure*}

The signal is modeled using a sum of a double Crystal Ball~\cite{CrystalBall,DoubleCrystalBall} and a Gaussian pdf. The power-law tails of the former component capture the outlying events in the resonance, while the additional Gaussian improves the description of the core shape, particularly at low masses.
The signal model is shown in Fig.~\ref{fig:sigModel} for two mass hypotheses, representing the lower and higher mass ranges. Assuming a linear dependence of the eight free parameters with respect to the diphoton mass, a parametric model starting from a nominal mass value of 40\GeV is constructed spanning the mass range in steps of 5\GeV. The model allows the extraction of the mass shape for each mass value. Given the high granularity of the mass sampling, a linear fit provides a reliable determination of each parameter at every mass value. 

The product of efficiency and acceptance for events entering the SR is evaluated using reference simulated signal samples with a spline interpolation across the mass range 10--70\GeV. 
The efficiency times acceptance curve as a function of the mass for events in the SR is mostly linear with some variation observed above 50\GeV, beyond which the trigger also enables the selection of less boosted topologies, as previously discussed. 

\begin{figure*}[!hbt]
\centering
\includegraphics[width=0.7\textwidth]{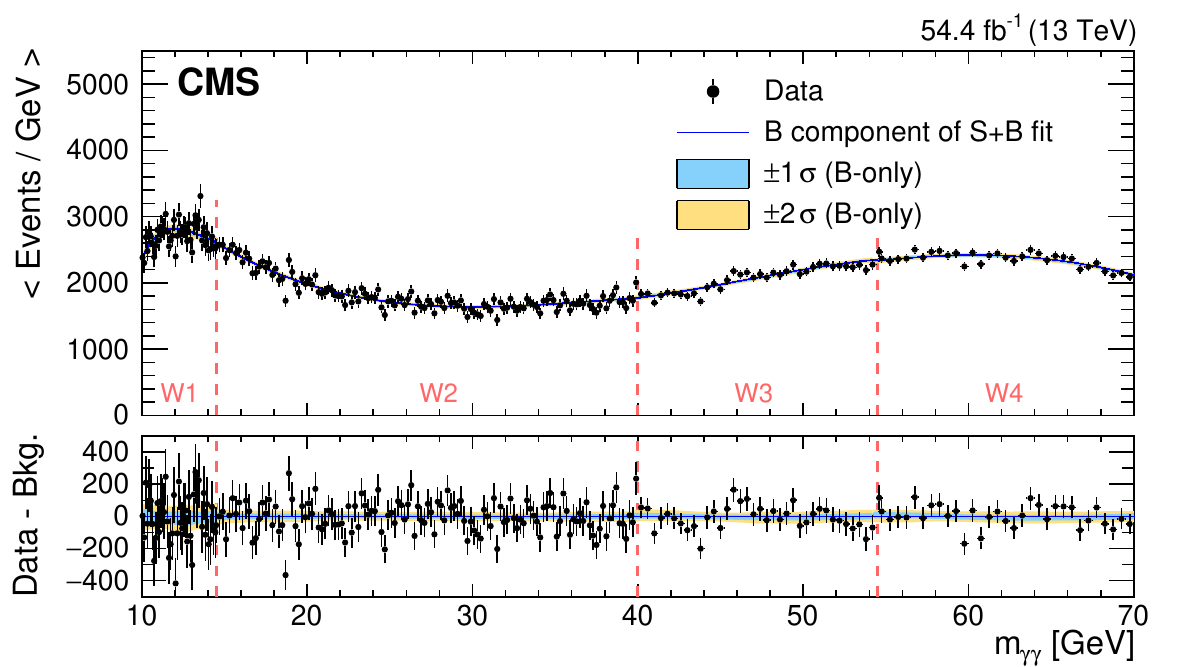}
\caption{Upper panel: The data (filled points) and background (dashed line) component of the signal-plus-background model are shown for each of the four windows (W1--W4) spanning the diphoton invariant mass spectrum by using the full data set from 10 to 70\GeV in the SR. The background fits include one (light blue) and two (yellow) standard deviation ($\upsigma$) uncertainties. Lower panel: Residuals in data and the uncertainty bands after subtracting the background fit. To illustrate the continuity of the data and fit models across the mass spectrum, event counts in both panels are scaled by the bin width of each window, leading to an average of event frequency computed for each bin throughout the mass spectrum.
}
\label{fig:bkgModel}
\end{figure*}

A dedicated strategy is developed to model the diphoton background mass spectrum in the range between 10 and 70\GeV, addressing the unique challenges of the background shape at very low masses. The spectrum shown in Fig.~\ref{fig:bkgModel} exhibits two distinct turn-on features: one around 12\GeV, arising from the photon isolation criteria starting to be inefficient for close-by objects from highly boosted diphoton candidates, and another near 50\GeV, driven by the photon \pt trigger thresholds. 
Additionally, the impact of the NN selection, which preferentially targets very boosted topologies, is responsible for the drop after the peak in the turn-on region at low mass. 
To address these complexities, the full mass range is divided into smaller subranges which can be described by a simple function.
In particular, the mass spectrum is divided into four subranges within the 10--70\GeV mass 
range with boundaries at 10.0, 14.5, 40.0, 54.5, and 70.0\GeV. For the background modeling, overlapping mass windows are used to ensure a stable description across the spectrum, 
while the non-overlapping intervals defined by the boundaries are adopted for the final statistical interpretation. Variable bin sizes are employed: 0.1, 0.2, 0.4, and 0.5\GeV from small to high masses, to account for the varying mass resolution of the signal resonance at different mass values.

Bernstein polynomial, exponential, and power-law functions are tested for modeling the background in each subrange. The minimum order needed for each function family to fit the data is determined using goodness-of-fit tests~\cite{goodnessoffit}, while the maximum order needed is determined using F-tests~\cite{ftest}. The minimum goodness-of-fit requirement is a p-value larger than 0.05, and the maximum order is determined when the p-value for the F-test becomes less than 0.05.
The selected functions in each window are then considered in the discrete profiling method~\cite{Dauncey:2014xga}. This technique estimates the systematic uncertainty associated with the background model by treating the background model choice as a discrete nuisance parameter in the likelihood fit to the data.
For this analysis, the set of candidate functions includes Bernstein polynomials from second up to fourth order, exponential functions with the exponent given by a polynomial up to third order, and power-law functions parameterized up to third order.
Figure~\ref{fig:bkgModel} shows the diphoton invariant mass spectrum obtained from 2018 data in the SR, split into four windows (labeled W1--W4), along with the corresponding set of best fit background functions selected in each of the subranges.

The signal and background modeling are incorporated into the final fitting strategy, which involves a binned maximum-likelihood fit to the diphoton mass spectrum to search for a narrow resonance. The mass range is scanned in fine steps, guided by the typical experimental resolution, to systematically test all mass hypotheses. 
Systematic uncertainties, described in the following Section, are also included as nuisance parameters in the likelihood.

\section{Systematic uncertainties}
\label{sec6:syst}
The largest uncertainties in this analysis arise from statistical uncertainties and background modeling. Various sources of experimental and theoretical uncertainties arising from detector effects and methodology are taken into account. Uncertainties related to the shape of the photon ID MVA distribution and the photon energy resolution, as detailed in Refs.~\cite{HIG-13-001,CMS-HIG-16-040}, are included.
The uncertainty in the photon ID MVA is derived from the chained quantile regression corrections of shower shape and isolation, and it amounts to about 3--4\%. Photon energy scale and resolution uncertainties are assessed using $\PZ\to\EE$ events and have a similar uncertainty of about 3--4\%. 
These two sources of uncertainty, along with that associated with the preselection scale factors (SFs), are the largest contributions. The preselection efficiency SFs, defined as the ratio between efficiencies in the data and in the simulation, are also determined using $\PZ\to\EE$ events.
The systematic uncertainty on the extrapolation from the kinematic phase space of the $\PZ$ peak to that of a lower mass resonance was assessed by examining events across different diphoton \pt bins, up to a diphoton \pt larger than 70\GeV. The data and corrected simulation were found to be compatible within the total uncertainty across the full range.
An integrated luminosity uncertainty of 2.5\% for the 2018 data taking~\cite{CMSlumi2018} is applied to the signal yield, along with an additional uncertainty of less than 1\% to account for mismodeling of the vertex multiplicity in simulation. 
The trigger efficiency uncertainty is measured using the tag-and-probe method with $\PZ\to\EE$ events and gives rise to efficiency variations of less than 0.5\%. Preselection, loose MVA ID, and electron veto SF uncertainties give variations in the expected number of signal events of about 5.0\%, 0.2\%, and 0.5\%, respectively.

Additional systematic uncertainties arise from several effects, as discussed in Refs.~\cite{HIG-13-001,CMS-HIG-16-040}: 
\begin{itemize}
    \item the modeling of the material budget upstream of the calorimeter (which can further modify the shower shapes in ECAL of photons compared to electrons),
    \item longitudinal nonuniformity in the light collection (which can affect the energy scale by modifying the shower profile in the crystal between electrons and photons),
    \item non-closure of photon energy scale corrections as a function of \pt,
    \item inaccuracies in the electromagnetic shower simulation, and
    \item vertex identification.
\end{itemize}

These uncertainties are propagated to the parametric signal model as shape uncertainties, leading to variations in the diphoton efficiency, mass scale, and resolution. Their overall effect is comparatively small with respect to the other systematic sources discussed above.

A set of theoretical uncertainties affecting both the signal acceptance and the selection efficiency, and hence the overall signal normalization, is also considered.
Renormalization and factorization uncertainties arise from missing higher-order terms in the perturbative calculation of the cross section. These uncertainties are evaluated by reweighting MC events with variations of the renormalization and factorization scales up and down by factors of two, where each variation is applied independently.
Their impact on the sensitivity of the limits is found to be at most 7\%.
The uncertainty in the strong coupling constant results in a 3\% systematic effect.
The uncertainty arising from the imperfect knowledge of the proton PDFs is evaluated by reweighting events with alternative sets of weights, following the \textsc{PDF4LHC} prescription~\cite{Butterworth:2015oua}, and is found to be negligible.

\section{Results}
\label{sec7:results}
\begin{figure*}[!hbt]
\centering
\includegraphics[width=0.7\textwidth]{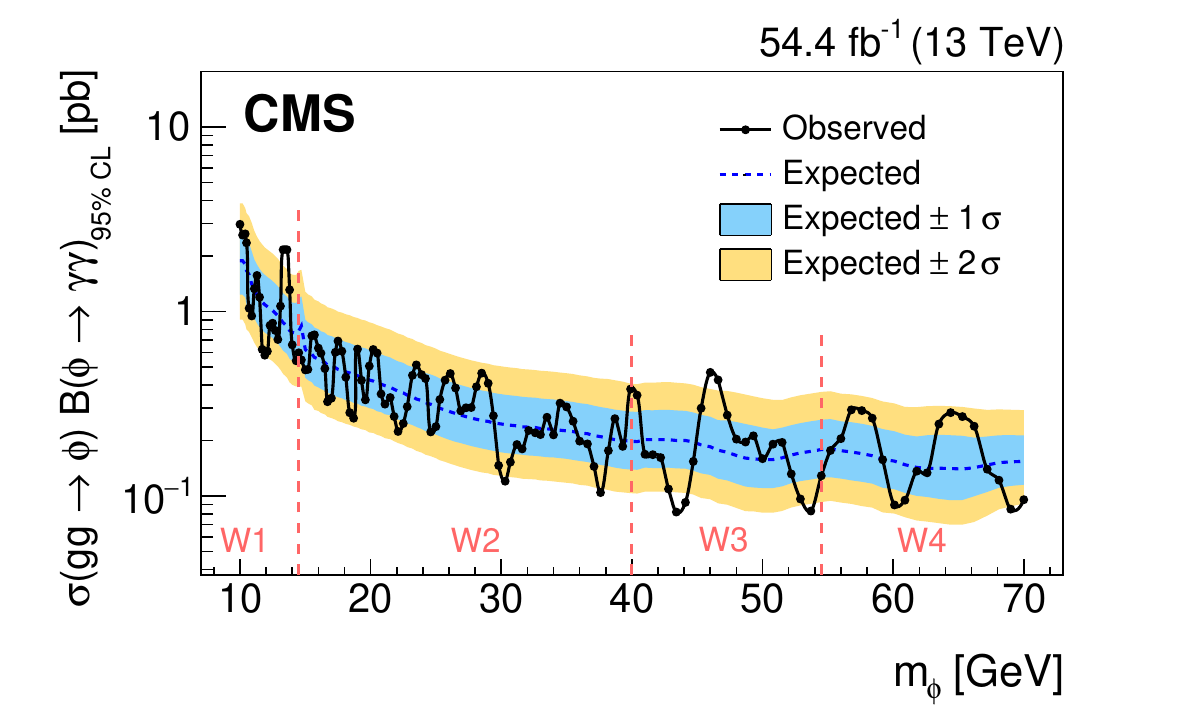}
\caption{Expected (dashed blue) and observed (solid black) limits at 95\% \CL on the gluon-fusion production cross section times the diphoton branching fraction $\upsigma ({\ggPHI}) \mathcal{B}(\PHIgg)$ of a narrow diphoton resonance, as functions of the mass in the range 10--70\GeV with minimal model-dependent assumptions, along with $1\,\upsigma$ (blue) and $2\,\upsigma$ (yellow) uncertainty bands from the expected limits. The four subranges spanning the diphoton spectrum, as employed for the background modeling, are labeled W1--W4.}
\label{fig:expObsLimits}
\end{figure*}

\begin{figure}[!hbt]
\centering
\includegraphics[width=\cmsFigWidth]{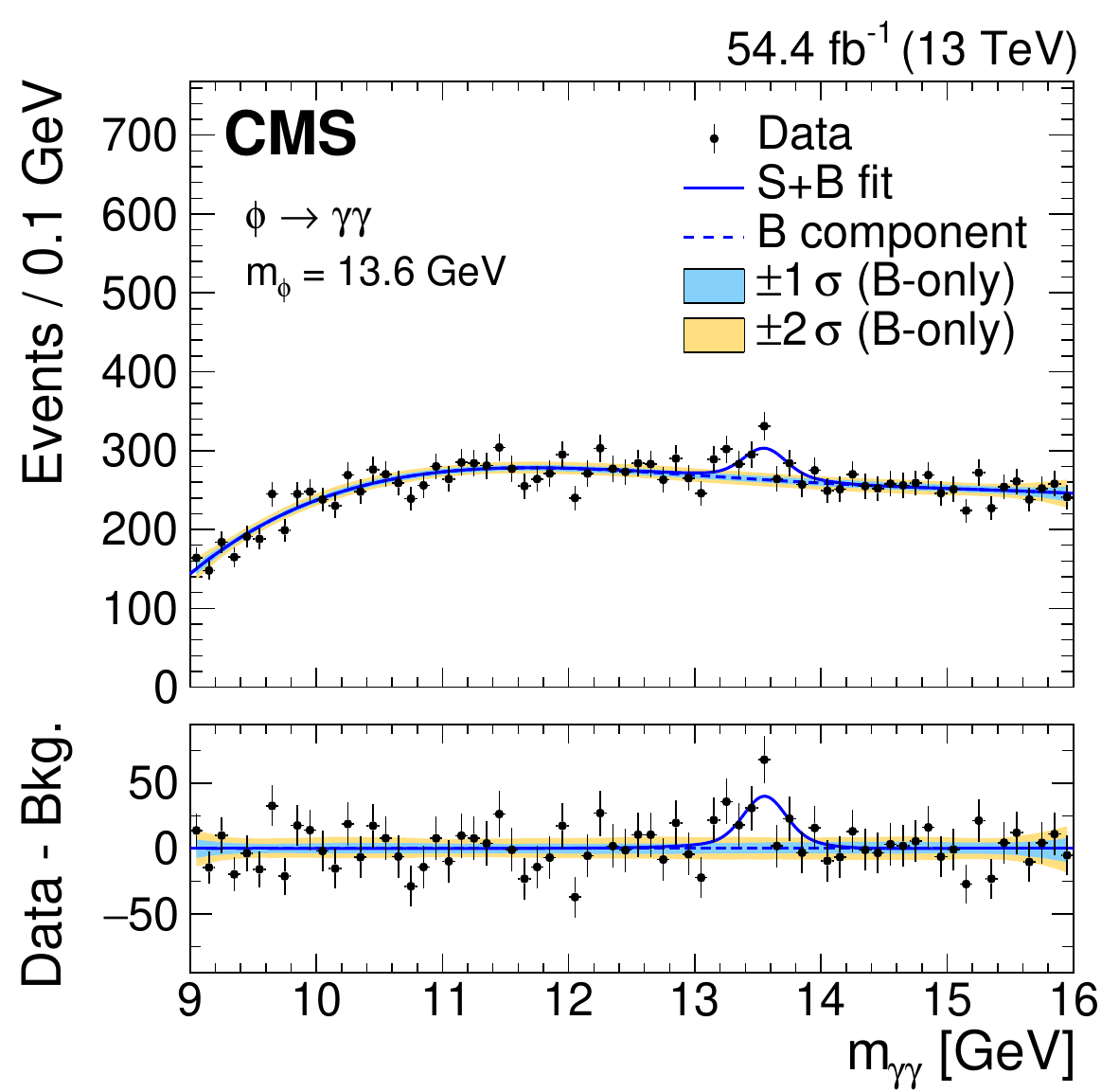}
\caption{Signal plus background fits (dark blue) are shown for the most significant excess in the SR at a mass hypothesis of 13.6\GeV, with signal yield freely floating. In the ratio pad, the background model is subtracted from the data. Uncertainty bands at ${\pm}1\,\upsigma$ (light blue) and ${\pm}2\,\upsigma$ (yellow) are obtained from pseudo-experiments generated separately for each window.}
\label{fig:mostSigExcess}
\end{figure}

A binned maximum likelihood fit to the diphoton invariant mass distribution across the mass range of 10--70\GeV is performed for events in the SR with the highest sensitivity, to derive limits on the production cross section times branching fraction of a prompt scalar resonance decaying to photon pairs.  
The systematic uncertainties discussed in the previous paragraph are incorporated into the likelihood through nuisance parameters.
The statistical analysis is performed using the profile likelihood approach in the {\textsc{Combine}} statistical tool~\cite{CMS:2024onh} to extract confidence intervals on the parameter of interest. 
Figure~\ref{fig:expObsLimits} presents the expected and observed limits at 95\% confidence level (\CL)~\cite{CLS1,CLS2} on the gluon-fusion production cross section times branching fraction to diphotons for a narrow scalar resonance. The variable granularity chosen for the computation of the observed and expected limit reflects the experimental resolution of the signal resonance across different mass ranges. The largest excess is observed at a mass hypothesis of 13.6\GeV, as presented in Fig.~\ref{fig:mostSigExcess}, corresponding to a local significance~\cite{Gross:2010qma} of 3.5 standard deviations. The calculation of the global significance~\cite{Gross:2010qma} is performed by using pseudo-experiments, and yields a result of ($1.9\pm0.1$) standard deviations. Five hundred Asimov data sets~\cite{Cowan:2010js} are generated as individual experiments, and the number of experiments with an excess larger than the most significant observed excess is counted. The ratio of these occurrences to the total number of pseudo-experiments gives the global probability, with the associated uncertainty computed as a binomial uncertainty.

Consistency with the results obtained by the CMS diphoton search in the mass range between 70 and 110\GeV~\cite{CMS:2024yhz} has been verified at 70\GeV.

A similar search has been performed by the ATLAS Collaboration using the full data set collected during 2016--2018~\cite{ATLAS-CONF-2022-018}, reporting limits on the fiducial production cross section times branching ratio to diphotons for a scalar resonance. As a rough comparison, the ATLAS result is extrapolated to the inclusive cross section using the fiducial cross section together with a mass-dependent parameterization of the fiducial acceptance at particle level, as provided in the corresponding HEPData record~\cite{hepdata.131600}. The CMS 2018 search is found to yield competitive sensitivities per unit of integrated luminosity in the low-mass region, and results that are already comparable to ATLAS in the higher mass region.

\subsection*{Interpretation in the context of the ALP model}
Upper limits on the production cross section of a hypothetical scalar diphoton resonance, as presented, are derived from the study of inclusive production with minimal model-dependent assumptions. These limits are therefore well-suited for reinterpretation within the ALP framework as constraints on the decay constant $f_{\parta}$, which governs the ALP ($\parta$) coupling to SM fields. In the scenario where the production is dominated by gluon fusion via the top quark loop depicted in Fig.~\ref{fig:diagram}, the ALP width is primarily determined by its effective coupling to gluons.
Effective field theories (EFTs) for ALPs~\cite{Brivio:2017ije} describe how the spontaneous breaking of a global $U$(1) symmetry, possibly a subgroup of a larger global symmetry $\mathcal{G}$, leads to the emergence of a massless \textit{CP}-odd Nambu--Goldstone boson (NGB) in the low-energy spectrum, called axion.
If the $U$(1) symmetry is only approximate, the NGB acquires a mass ($m_{\parta}$), becoming a pseudo-NGB, commonly referred to as an ALP.
As detailed in Ref.~\cite{Brivio:2017ije}, within the framework of linear realizations of EW symmetry breaking, the most general linear bosonic Lagrangian for an ALP ($\mathcal{L}_{\text{eff}}^{\text{linear}}$) with axial couplings to SM gauge bosons can be written as:

\ifthenelse{\boolean{cms@external}}
{ 
\begin{multline*}
\mathcal{L}_{\text{eff}}^{\text{linear}} = \mathcal{L}^{\text{LO}} + \delta\mathcal{L}_{\parta}^{\text{bosonic}} = \mathcal{L}^{\text{SM}} \\
+\frac{1}{2}(\delta_{\mu}\parta) (\delta^{\mu}\parta) + c_{\tilde{B}}A_{\tilde{B}} + c_{\tilde{W}}A_{\tilde{W}} + c_{\tilde{G}}A_{\tilde{G}} + c_{\parta\Phi}O_{\parta\Phi}.
\end{multline*}
}
{
\begin{equation*}
\mathcal{L}_{\text{eff}}^{\text{linear}} = \mathcal{L}^{\text{LO}} + \delta\mathcal{L}_{\parta}^{\text{bosonic}} = \mathcal{L}^{\text{SM}} + \frac{1}{2}(\delta_{\mu}\parta) (\delta^{\mu}\parta) + c_{\tilde{B}}A_{\tilde{B}} + c_{\tilde{W}}A_{\tilde{W}} + c_{\tilde{G}}A_{\tilde{G}} + c_{\parta\Phi}O_{\parta\Phi}.
\end{equation*}
}
The LO Lagrangian ($\mathcal{L}^{\text{LO}}$) is the combination of the SM one ($\mathcal{L}^{\text{SM}}$) and the ALP kinetic term. The bosonic part incorporates the NLO corrections involving the field $\parta$ and describes the interactions with the B, W, and G fields. These fields represent the electromagnetic, EW, and QCD field strength tensors, respectively, with the coefficients $c_{\tilde{i}}$ representing the anomalies of the global $U$(1) with $SU$(2)$\times U(1)_{Y}$ and $SU$(3), respectively.
Denoting by $f_{\parta}$ the scale associated to the physics of the ALP, its field $\parta$ enters in effective operators with a scaling effect by powers of $\parta/f_{\parta}$. Each operator can be expressed as:
\ifthenelse{\boolean{cms@external}}{
\begin{align*}
A_{\tilde{B}} &= - B_{\mu\nu} \tilde{B}^{\mu\nu} \frac{\parta}{f_{\parta}},&
A_{\tilde{W}} &= - W_{\mu\nu} \tilde{W}^{\mu\nu} \frac{\parta}{f_{\parta}}, \\
A_{\tilde{G}} &= - G_{\mu\nu} \tilde{G}^{\mu\nu} \frac{\parta}{f_{\parta}},&
\textrm{O}_{\parta\Phi} &= i(\Phi^{\dagger}\overset{\leftrightarrow}{D}_{\mu}\Phi) \frac{\delta^{\mu}\parta}{f_{\parta}}.
\end{align*}
}
{
\begin{equation*}
A_{\tilde{B}} = - B_{\mu\nu} \tilde{B}^{\mu\nu} \frac{\parta}{f_{\parta}},
A_{\tilde{W}} = - W_{\mu\nu} \tilde{W}^{\mu\nu} \frac{\parta}{f_{\parta}},
A_{\tilde{G}} = - G_{\mu\nu} \tilde{G}^{\mu\nu} \frac{\parta}{f_{\parta}},
\textrm{O}_{\parta\Phi} = i(\Phi^{\dagger}\overset{\leftrightarrow}{D}_{\mu}\Phi) \frac{\delta^{\mu}\parta}{f_{\parta}}.
\end{equation*}
}

We note that $\textrm{O}_{\parta\Phi}$ represents an additional contribution to the longitudinal component of the EW gauge fields, which is neglected here. Therefore, $c_{\parta\Phi}$ is set to zero. 

The ALP model is implemented in \textsc{FeynRules}~\cite{Alloul:2013bka}, according to the theoretical framework presented in Ref.~\cite{Brivio:2017ije}. 
The \MGvATNLO v2.6.5 matrix element generator~\cite{AMCAT} is used to extract theoretical cross sections and branching fractions of ALP production expected at the LHC. 
The LO process $\Pg\Pg\to \parta \to \PGg \PGg$ is generated along with the associated production process $\Pg\Pg\to \parta\Pj \to \PGg \PGg \Pj$, where \Pj denotes an additional jet.
No transverse momentum selection is applied to the additional jet in the generation of signal events. A detailed study of jet merging within \PYTHIA has been conducted, focusing on tuning the merging scale through the MLM matching scheme~\cite{Alwall:2007fs}. The merging scale parameters, \texttt{xqcut} at the \MGvATNLO level and \texttt{qcut} at the \PYTHIA level, are optimized by setting \texttt{qcut} to approximately twice \texttt{xqcut}. 
In this study, \texttt{xqcut} is set to 10\GeV and \texttt{qcut} to 20\GeV.
This configuration provides a good description of the transition between different jet multiplicities across the entire mass range.
The set of PDF used is \textsc{NNPDF31\_nnlo\_as\_0118\_nf\_4\_mc\_hessian}~\cite{Ball:2017nwa}. 
To account for NLO corrections, a constant K factor of 2 is applied to these cross sections, as derived in Refs.~\cite{Mariotti:2017vtv,Gershtein:2020mwi} for a similar topology. Since the study presented in this paper focuses on the inclusive gluon-fusion production, the theoretical cross sections are computed without any acceptance requirement on the photons.
The extrapolation of the limits on the production cross section of a scalar resonance to the ALP phase space is performed under the assumption that the acceptance is the same at NLO for the ALP production as for a SM-like $\PHI$ boson with the same mass produced in gluon fusion. To validate this assumption, the number of signal events passing the full SR selection for the ALP model generated at LO was compared to that of the SM-like $\PHI$ boson generated at NLO. The difference was at most 13\% for the 10\GeV mass hypothesis, well within the nominal systematic uncertainties associated with the signal selection.

The value of $f_{\parta}$ represents the cutoff scale of the EFTs, which is naturally expected to be large. 
When considering only the coupling to photons, the ALP decay width ($\PGg_{\parta}$) can be written as

\begin{equation*}
\PGg_{\parta} = \frac{m_{\parta}^3 (c_{\tilde{B}} c_{\theta}^2 + c_{\tilde{W}} s_{\theta}^2)^2}{4 \pi f_{\parta}^2},
\end{equation*}

with $c_{\theta}$ and $s_{\theta}$ denoting the cosine and sine of the weak mixing angle.
However, for the branching fraction, the dependence on $f_{\parta}$ cancels out in the ratio between the partial and the total widths. Thus, the ALP production cross section times branching fraction, ${\sigma(\Pg\Pg\to \parta)}{\mathcal{B}(\parta \to\PGg\PGg)}$, is proportional to $(\frac{1}{f_{\parta}})^2$.

\begin{figure*}[!htb]
\centering
\includegraphics[width=0.7\textwidth]{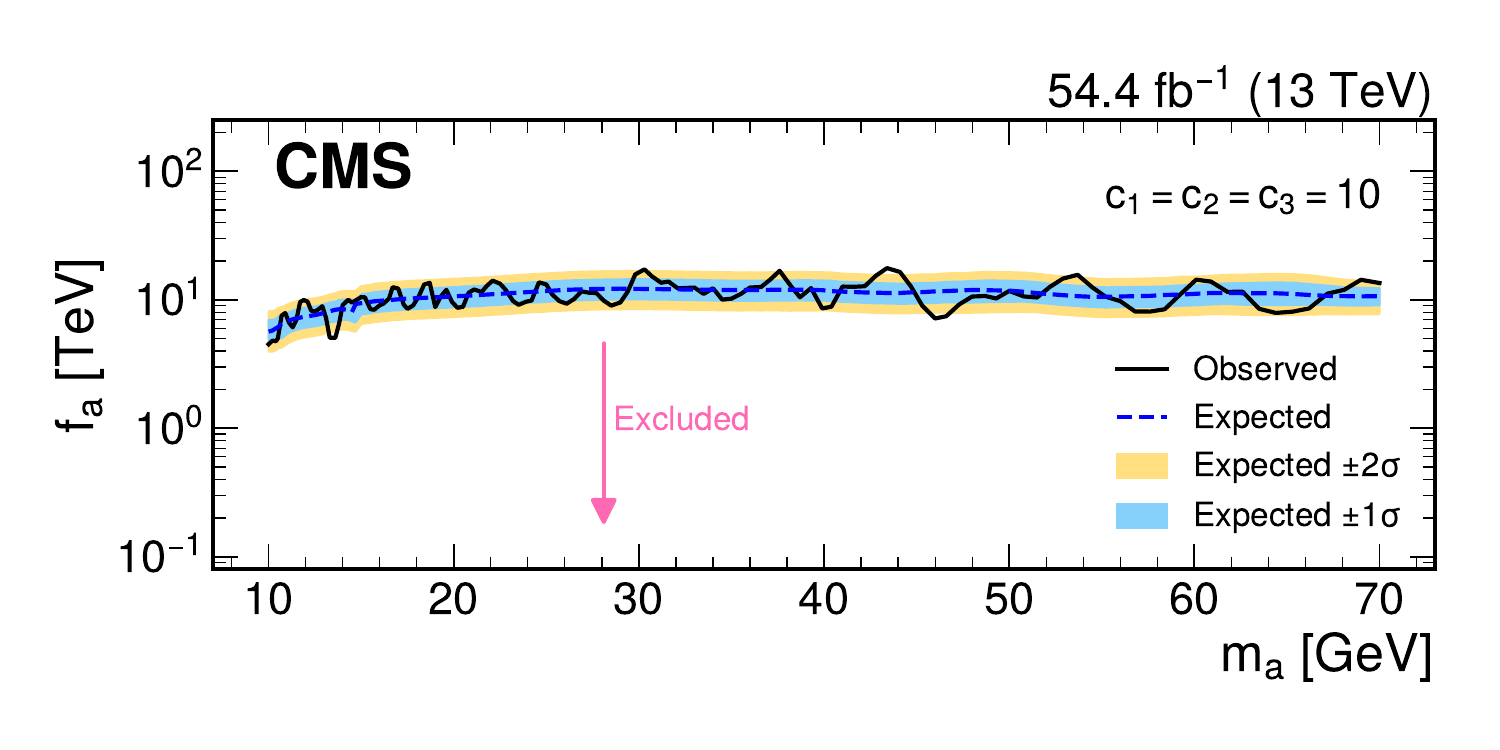}
\caption{Recasting of the expected (dashed blue line) and observed (solid black line) upper limits on the gluon-fusion production cross section of a diphoton resonance as a function of its mass in the range 10--70\GeV into the ALP parameter space as constraints on the decay constant $f_{\parta}$, assuming $c_1 = c_2 = c_3 = 10$. The upper limit on the production cross section corresponds to a lower bound on $f_{\parta}$. 
}
\label{fig:alpLimit}
\end{figure*}

The interpretation of the limits in the ALP parameter space as constraints on the decay constant $f_{\parta}$ as a function of the diphoton mass is shown in Fig.~\ref{fig:alpLimit}, assuming $c_1=c_2=c_3=10$, where $i=1,2,3$ correspond to the couplings to the hypercharge, weak isospin, and QCD gauge fields, respectively. 
The rescaling between the coefficients $c_i$ and $c_{\tilde{i}}$ accounts for their dependence on the electromagnetic coupling $\upalpha_\mathrm{EM}$, the Weinberg angle, and the strong coupling constant $\upalpha_\textrm{S}$. The corresponding relations can be expressed as:

\begin{equation*}
c_{\tilde{B}} = \frac{5}{3}\frac{\upalpha_{\mathrm{EM}}}{c_{\theta}^2} \frac{c_1} {4 \pi}, \hspace{0.4cm} 
c_{\tilde{W}} = \frac{\upalpha_{\mathrm{EM}}}{s_{\theta}^2} \frac{c_2} {4 \pi}, \hspace{0.4cm}
c_{\tilde{G}} = \upalpha_{\textrm{S}} \frac{c_3} {4 \pi}.
\end{equation*}
The upper limit on the production cross section corresponds to a lower bound on $f_{\parta}$, since larger values of $f_{\parta}$ lead to smaller cross sections, which are currently beyond our sensitivity. 
The benchmark value $c_i=10$ is chosen for comparison with other experiments, and the results can be reinterpreted for different choices of the anomaly coefficients and decay constant. The assumption of equal magnitudes for the anomaly coefficients $c_i$ is motivated by gauge coupling unification in Grand Unified Theory scenarios.    

\section{Summary}
\label{sec8:summary}
A novel search for a narrow-width resonance has been presented in the diphoton invariant mass range of 10--70\GeV, enabled for the first time by a new trigger introduced for the CMS data taking in 2018. An integrated luminosity corresponding to 54.4\fbinv of proton-proton collision data at a center-of-mass energy of 13\TeV is analyzed. 
The gluon-fusion production cross section times diphoton branching fraction of a narrow and prompt resonance is studied across a broad mass range with minimal model-dependent assumptions.
No significant diphoton signal is observed. The 95\% confidence level upper limit is set on the production cross section times diphoton branching fraction, as a function of the resonance mass. The main sources of uncertainty stem from the limited number of collected data and from background modeling. 
Finally, an interpretation of the limits within an effective field theory framework for axion-like particles is also provided, setting lower limits on its decay constant parameter within 4--15\TeV, mildly dependent on the mass of the particle.

\begin{acknowledgments}
We congratulate our colleagues in the CERN accelerator departments for the excellent performance of the LHC and thank the technical and administrative staffs at CERN and at other CMS institutes for their contributions to the success of the CMS effort. In addition, we gratefully acknowledge the computing centers and personnel of the Worldwide LHC Computing Grid and other centers for delivering so effectively the computing infrastructure essential to our analyses. Finally, we acknowledge the enduring support for the construction and operation of the LHC, the CMS detector, and the supporting computing infrastructure provided by the following funding agencies: SC (Armenia), BMBWF and FWF (Austria); FNRS and FWO (Belgium); CNPq, CAPES, FAPERJ, FAPERGS, and FAPESP (Brazil); MES and BNSF (Bulgaria); CERN; CAS, MoST, and NSFC (China); MINCIENCIAS (Colombia); MSES and CSF (Croatia); RIF (Cyprus); SENESCYT (Ecuador); ERC PRG and PSG, TARISTU24-TK10 and MoER TK202 (Estonia); Academy of Finland, MEC, and HIP (Finland); CEA and CNRS/IN2P3 (France); SRNSF (Georgia); BMFTR, DFG, and HGF (Germany); GSRI (Greece); NKFIH (Hungary); DAE and DST (India); IPM (Iran); SFI (Ireland); INFN (Italy); MSIT and NRF (Republic of Korea); MES (Latvia); LMTLT (Lithuania); MOE and UM (Malaysia); BUAP, CINVESTAV, CONACYT, LNS, SEP, and UASLP-FAI (Mexico); MOS (Montenegro); MBIE (New Zealand); PAEC (Pakistan); MES, NSC, and NAWA (Poland); FCT (Portugal); MESTD (Serbia); MICIU/AEI and PCTI (Spain); MOSTR (Sri Lanka); Swiss Funding Agencies (Switzerland); MST (Taipei); MHESI (Thailand); TUBITAK and TENMAK (T\"{u}rkiye); NASU (Ukraine); STFC (United Kingdom); DOE and NSF (USA).

\hyphenation{Rachada-pisek} Individuals have received support from the Marie-Curie program and the European Research Council and Horizon 2020 Grant, contract Nos.\ 675440, 724704, 752730, 758316, 765710, 824093, 101115353, 101002207, 101001205, and COST Action CA16108 (European Union); the Leventis Foundation; the Alfred P.\ Sloan Foundation; the Alexander von Humboldt Foundation; the Science Committee, project no. 22rl-037 (Armenia); the Fonds pour la Formation \`a la Recherche dans l'Industrie et dans l'Agriculture (FRIA) and Fonds voor Wetenschappelijk Onderzoek contract No. 1228724N (Belgium); the Beijing Municipal Science \& Technology Commission, No. Z191100007219010, the Fundamental Research Funds for the Central Universities, the Ministry of Science and Technology of China under Grant No. 2023YFA1605804, the Natural Science Foundation of China under Grant No. 12535004, and USTC Research Funds of the Double First-Class Initiative No.\ YD2030002017 (China); the Ministry of Education, Youth and Sports (MEYS) of the Czech Republic; the Shota Rustaveli National Science Foundation, grant FR-22-985 (Georgia); the Deutsche Forschungsgemeinschaft (DFG), among others, under Germany's Excellence Strategy -- EXC 2121 ``Quantum Universe" -- 390833306, and under project number 400140256 - GRK2497; the Hellenic Foundation for Research and Innovation (HFRI), Project Number 2288 (Greece); the Hungarian Academy of Sciences, the New National Excellence Program - \'UNKP, the NKFIH research grants K 131991, K 133046, K 138136, K 143460, K 143477, K 146913, K 146914, K 147048, 2020-2.2.1-ED-2021-00181, TKP2021-NKTA-64, and 2025-1.1.5-NEMZ\_KI-2025-00004 (Hungary); the Council of Science and Industrial Research, India; ICSC -- National Research Center for High Performance Computing, Big Data and Quantum Computing, FAIR -- Future Artificial Intelligence Research, and CUP I53D23001070006 (Mission 4 Component 1), funded by the NextGenerationEU program (Italy); the Latvian Council of Science; the Ministry of Education and Science, project no. 2022/WK/14, and the National Science Center, contracts Opus 2021/41/B/ST2/01369, 2021/43/B/ST2/01552, 2023/49/B/ST2/03273, and the NAWA contract BPN/PPO/2021/1/00011 (Poland); the Funda\c{c}\~ao para a Ci\^encia e a Tecnologia (Portugal); the National Priorities Research Program by Qatar National Research Fund; MICIU/AEI/10.13039/501100011033, ERDF/EU, ``European Union NextGenerationEU/PRTR", and Programa Severo Ochoa del Principado de Asturias (Spain); the Chulalongkorn Academic into Its 2nd Century Project Advancement Project, the National Science, Research and Innovation Fund program IND\_FF\_68\_369\_2300\_097, and the Program Management Unit for Human Resources \& Institutional Development, Research and Innovation, grant B39G680009 (Thailand); the Eric \& Wendy Schmidt Fund for Strategic Innovation through the CERN Next Generation Triggers project under grant agreement number SIF-2023-004; the Kavli Foundation; the Nvidia Corporation; the SuperMicro Corporation; the Welch Foundation, contract C-1845; and the Weston Havens Foundation (USA).
\end{acknowledgments}

\bibliography{auto_generated} 
\cleardoublepage \appendix\section{The CMS Collaboration \label{app:collab}}\begin{sloppypar}\hyphenpenalty=5000\widowpenalty=500\clubpenalty=5000\cmsinstitute{Yerevan Physics Institute, Yerevan, Armenia}
{\tolerance=6000
A.~Hayrapetyan, V.~Makarenko\cmsorcid{0000-0002-8406-8605}, A.~Tumasyan\cmsAuthorMark{1}\cmsorcid{0009-0000-0684-6742}
\par}
\cmsinstitute{Institut f\"{u}r Hochenergiephysik, Vienna, Austria}
{\tolerance=6000
W.~Adam\cmsorcid{0000-0001-9099-4341}, J.W.~Andrejkovic, L.~Benato\cmsorcid{0000-0001-5135-7489}, T.~Bergauer\cmsorcid{0000-0002-5786-0293}, M.~Dragicevic\cmsorcid{0000-0003-1967-6783}, C.~Giordano\cmsorcid{0000-0001-6317-2481}, P.S.~Hussain\cmsorcid{0000-0002-4825-5278}, M.~Jeitler\cmsAuthorMark{2}\cmsorcid{0000-0002-5141-9560}, N.~Krammer\cmsorcid{0000-0002-0548-0985}, A.~Li\cmsorcid{0000-0002-4547-116X}, D.~Liko\cmsorcid{0000-0002-3380-473X}, M.~Matthewman, I.~Mikulec\cmsorcid{0000-0003-0385-2746}, J.~Schieck\cmsAuthorMark{2}\cmsorcid{0000-0002-1058-8093}, R.~Sch\"{o}fbeck\cmsAuthorMark{2}\cmsorcid{0000-0002-2332-8784}, D.~Schwarz\cmsorcid{0000-0002-3821-7331}, M.~Shooshtari\cmsorcid{0009-0004-8882-4887}, M.~Sonawane\cmsorcid{0000-0003-0510-7010}, W.~Waltenberger\cmsorcid{0000-0002-6215-7228}, C.-E.~Wulz\cmsAuthorMark{2}\cmsorcid{0000-0001-9226-5812}
% ...
\par}
\vskip\cmsinstskip
\dag:~Deceased\\
$^{1}$Also at Yerevan State University, Yerevan, Armenia\\
$^{2}$Also at TU Wien, Vienna, Austria\\
% ...
\end{sloppypar}
\end{document}